# Novel Weight Update Scheme for Hardware Neural Network based on Synaptic Devices Having Abrupt LTP or LTD Characteristics

Junmo Lee[1]*, Joon Hwang[1], Youngwoon Cho[2], Sangbum Kim[2], and Jongho Lee[1]

* wnsah3817@snu.ac.kr

[1]Department of Electrical and Computer Engineering, Seoul National University
[2]Department of Materials Science and Engineering, Seoul National University

**Abstract**

Mitigating nonlinear weight update characteristics is one of the main challenges in designing neural networks based on synaptic devices. This paper presents a novel weight update method named conditional reverse update scheme (CRUS) for hardware neural network (HNN) consisting of synaptic devices with highly nonlinear or abrupt conductance update characteristics. We formulate a linear optimization method of conductance in synaptic devices to reduce the average deviation of weight changes from those calculated by the Stochastic Gradient Rule (SGD) algorithm. We introduce a metric called update noise (UN) to analyze the training dynamics during training. We then design a weight update rule that reduces the UN averaged over the training process. The optimized network achieves >90% accuracy on the MNIST dataset under highly nonlinear long-term potentiation (LTP) and long-term depression (LTD) conditions while using inaccurate and infrequent conductance sensing. Furthermore, the proposed method shows better accuracy than previously reported nonlinear weight update mitigation techniques under the same hardware specifications and device conditions. It also exhibits robustness to temporal variations in conductance updates. We expect our scheme to relieve design requirements in device and circuit engineering and serve as a practical technique that can be applied to future HNNs.

## INTRODUCTION

Recently, hardware neural network (HNN) based on synaptic devices has received significant interest due to its potential to overcome the limitations of von Neuman architecture. In particular, their capability to tune and store the conductance makes synaptic devices well-suited for emulating biological behaviors such as long-term potentiation (LTP), long-term depression (LTD), and spike-timing-dependent plasticity (STDP) [1-2]. Such capabilities of synaptic devices can be utilized to build up a new paradigm of computation that can solve complex computational tasks consuming low power. Numerous types of synaptic devices that rely on various physical mechanisms have been explored in recent years [1,3]. It is well known that HNNs based on these synaptic devices have advantages in size, speed, and power consumption compared to those in conventional digital CMOS circuits [4-6]. Despite the bright prospects of synaptic device-based HNN, which is capable of parallel and energy-efficient computations, it has several issues that must be resolved before it can be implemented.

One of the main challenges hampering the practical application of synaptic devices is nonlinearity in conductance programming, which arises from the complex drift and diffusion mechanisms of the ions or vacancies [7-8]. For example, Phase Change Memory (PCM) and $Pr_{0.7}Ca_{0.3}MnO_3$ (PCMO) based synaptic devices exhibit abrupt LTD characteristics that make their conductance values rapidly fall to their minimum conductance states [9-11]. Due to this inherent non-linearity in the conductance modulation process, HNNs based on synaptic devices always suffer from a mismatch between the expected weight change and the actual weight change. While deep learning can have an inherent resilience to a moderate level of perturbation, excessive non-linearity in weight updates can be harmful to learning [12]. This problem has motivated many researchers to optimize the non-linearity of synaptic devices through various approaches. However, the linear optimization of devices changes their properties such as their on-off ratios, the number of possible conductance states, their endurance cycles, and the pulse-to-pulse variation, affecting the performance and cost of the hardware [13-14]. Moreover, to suppress the non-linearity using the pulse scheme, it is necessary to precisely control the duration and amplitude of the pulse considering the conductivity of the synaptic device [15]. There is also a physical limitation in the scaling of the pulse duration as the success probability of resistance switching decreases with pulse duration scaling [16-17]. Therefore, optimizing the non-linearity through device engineering or pulse schemes may not lead to performance enhancement at the system level [18]. From this point of view, algorithmic techniques to mitigate the non-linearity may be more appropriate than pulse shape tuning or device-level engineering in the design of HNNs.

In this work, we present a novel weight update scheme named conditional reverse update scheme (CRUS) to train HNNs composed of synaptic devices whose LTP and LTD characteristics are highly nonlinear or abrupt. To this end, the influence of update noise on neural networks is carefully analyzed through the introduction of an update noise evaluation metric. We believe that the methodology to analyze update noise and the proposed weight update scheme will be useful in optimizing the performance of HNN based on synaptic devices and can be applied to various synaptic devices.

## BIDIRECTIONAL FLASH-TYPE SYNAPTIC DEVICE

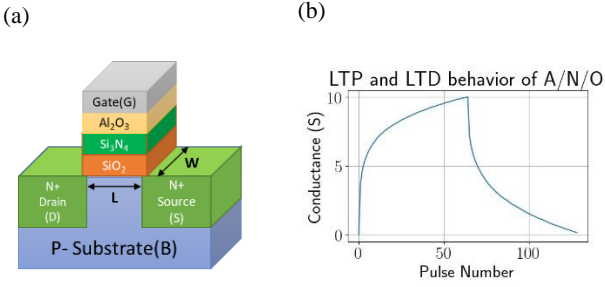

Fig. 1. (a) A schematic 3D view of A/N/O flash FET. (b) LTP and LTD behavior of A/N/O flash FET.

We present a flash-type memory device with abrupt and nonlinear conductance update characteristics. A schematic 3D view of this device is illustrated in Fig. 1(a). The structure of this device is identical to that of an *n*-channel MOSFET with a gate insulator stack of $Al_2O_3/Si_3N_4/SiO_2$. The thicknesses of $Al_2O_3$, $Si_3N_4$, and $SiO_2$ are 2 nm, 3 nm, and 9 nm, respectively. The W (Channel Width) is 50 µm and the L (Channel Length) is 0.8 µm. A/N/O device is capable of bidirectional weight (conductance) change through program (for LTD) or erase (for LTP) operations. During the program process, successive pulses are applied to the gate to trap charges in the $Si_3N_4$ layer. In contrast, during the erase process, the charges in the $Si_3N_4$ layer are detrapped by applying pulses to the gate. The magnitude of the threshold voltage shift, which is determined by the charge density stored in the $Si_3N_4$ layer, modulates the conductance of the synaptic device. Program and erase pulses we use are 5 V for 80 µs, and -6 V for 45 us, respectively. Here, the bias condition is $V_D=V_S=V_B=0$ V. Conductance values are read out by setting $V_{ds}$=0.1 V, $V_{gs}$=0.7 V. Fig. 1(b) shows the measured LTP and LTD behaviors of A/N/O device. The noticeable characteristic of the LTP curve is that the conductance changes are extremely abrupt in the early stage of pulse application. For the design of HNN based on this device, we develop a novel weight update method that can effectively mitigate the LTP and LTD characteristics that exhibit abruptness and high non-linearity.

## RELATED WORK

Recently, various effective non-linearity mitigation methods for HNN based on synaptic devices have been proposed. However, there are still several issues to be addressed.

First, the level of non-linearity tested in previous works may be inappropriate to simulate highly nonlinear or abrupt conductance update characteristics. In this work, we use the following equations to model the LTP and LTD curve [19]:

$$G_{LTP}(P) = B\left(1 - e^{\frac{P}{A}}\right) + G_{min} \quad (1)$$

$$G_{LTD}(P) = -B\left(1 - e^{\frac{P-P_{max}}{A}}\right) + G_{max} \quad (2)$$

$$B = \frac{G_{max} - G_{min}}{1 - e^{-\frac{P_{max}}{A}}} \quad (3)$$

Here, $G_{max}$ and $G_{min}$ are the maximum and minimum conductance of a synaptic device extracted from experimental data, respectively. A and B are the fitting parameters used to model the LTP and LTD curves. $P_{max}$ is the maximum number of available conductance steps of a synaptic device. The NL(LTP) (Non-linearity level of LTP) and NL(LTD) (Non-linearity level of LTD) range from -9 to 9 and these values are used to calculate the value of A. NL(LTP) and NL(LTD), and their corresponding LTP and LTD curves are illustrated in Fig. 2(a)-(b). We normalize the conductance value of synaptic devices to a range between 0 ($G_{min}$) and 10 ($G_{max}$) in the rest of this paper. To measure the magnitude of the abruptness of LTP or LTD characteristics, we define AF (Abruptness Factor) as follows:

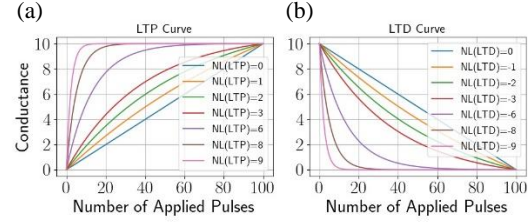

Fig. 2. (a) LTP and (b) LTD curve corresponding to different NL(LTP) and NL(LTD) values.

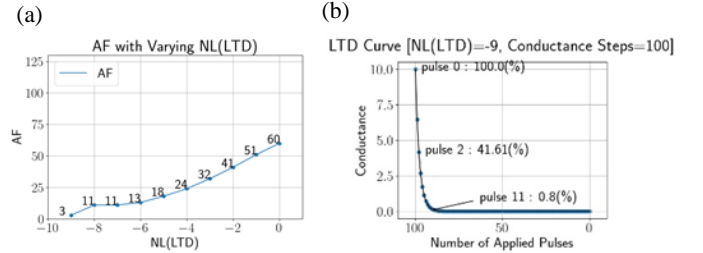

Fig. 3. (a) Abruptness Factor (LTD) with varying NL(LTD) (b) Abruptness of LTD curve at NL(LTD)=-9. (b) shows that the LTD curve with NL=-9 is extremely abrupt in the early stage of pulse application.

$$AF = \begin{cases} 100\frac{C_{LTP}}{P_{max}}, & for\ LTP \\ 100\frac{C_{LTD}}{P_{max}}, & for\ LTD \end{cases} \quad (4)$$

where $C_{LTP}$ represents the minimum number of pulses that should be applied to change the minimum conductance value by 60% of $(G_{max} - G_{min})$, $C_{LTD}$ represents the minimum number of pulses that should be applied to change the maximum conductance value by 60% of $(G_{max} - G_{min})$, and $P_{max}$ represents the maximum number of conductance steps in a synaptic device. AF can range from 0 - 60. Fig. 3(a) presents the NL and its corresponding AF. Note that there are cases when Eqs. (1)-(3) are not suitable to model the LTP or LTD curves of some synaptic devices [20]. In these cases, AF can be a simple and useful measure of the abruptness of LTP or LTD characteristics.

The piece-wise linear (PL) method is a recently introduced linear optimization technique [21]. In this method, the duration of the pulses for weight update is scaled by a scaling factor to compensate for nonlinear conductance changes. The scaling factor is determined after sensing the conductance of synaptic devices at every training iteration. This method was tested under NL ranging from -6 to 6 using the LTP and LTD curve modeled by Eqs. (1)-(3). In addition, a hardware-friendly stochastic and adaptive learning method has been proposed to counter the PL method [22]. However, it was also tested under NL ranging from approximately -6 to 6 and showed similar or lower accuracy than the PL method. Unfortunately, such NL values might be too low to capture the nonlinear or abrupt update characteristics of some synaptic devices. For example, the NL(LTD) of a reported PCMO device using the non-linearity model defined by Eqs. (1)-(3) is -6.76 [19]. The A/N/O device mentioned above has an AF of 9.375 for LTP and 17.1875 for LTD. Considering that the AF corresponding to NL(LTP) = 6 is 13, setting NL(LTP) = 6 is inappropriate to simulate abrupt LTP behavior in A/N/O device. More importantly, engineering the device structure to tune the LTP and LTD characteristics offsets crucial device properties that ensure high training performance in HNN based on synaptic devices [13,14,18]. This suggests that strategically increasing the non-linearity level or decreasing the AF might be an attractive option in synaptic device engineering. Therefore, devising a non-linearity mitigation technique that can work for high non-linearity levels and a low AF ($|NL(LTP)| \geq 8$ or $|NL(LTD)| \geq 8$, AF$\leq 11$) would be a promising approach in HNN design.

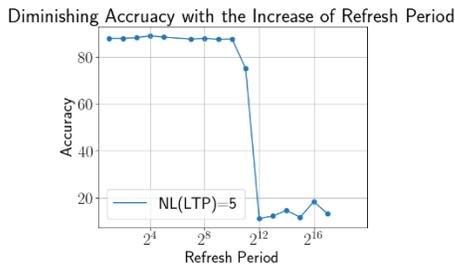

Fig. 4. Trade-off between accuracy and the refresh period under refresh scheme. The accuracy starts to decrease if the refresh period exceeds $2^{10}$.

Furthermore, non-linearity mitigation methods that require frequent and accurate conductance sensing may lead to high hardware costs. Some of the successful methods that guarantee acceptable accuracy under high NL settings require conductance sensing at every training iteration. They also sacrifice sensing precision to achieve ~90 % accuracy on the MNIST dataset under high NL settings [21,23]. Representatively, the refresh method was presented as an effective solution to yield high accuracy in HNN based on highly nonlinear devices such as PCM devices [24-25]. In this method, only LTP updates are used for weight updating. For this reason, the conductance values of synaptic devices are periodically sensed and rewritten into the devices to prevent the saturation of conductance. The key advantage of the refresh method is that if LTP is linear enough, high accuracies can be achieved regardless of the NL(LTD) value. Despite its effectiveness in boosting the performance and immunity to resistance drift effects [26], the periodical sensing and rewriting process induces a trade-off in the refresh period (number of training iterations between successive refresh processes), training power, and performance [9,27]. From Fig. 4, it can be seen that the accuracy rapidly drops above a certain refresh period in the case of the network presented in this paper. Thus, to create practical HNN, it may be more advantageous to design a weight update method that exhibits more robustness to infrequent and inaccurate conductance sensing and still achieves high accuracy (>90%).

Moreover, many studies on HNN based on synaptic devices lack quantitative analysis on how the non-linearity of LTP and LTD degrade the accuracy of the network [9,23,28]. Quantitative analysis on the impact caused by nonlinear weight update might help elucidate the origin of accuracy degradation.

Accordingly, our work proposes a hardware-efficient weight update rule and update noise evaluation metrics to optimize HNN based on synaptic devices. By designing a weight update rule that requires 1-bit precision and infrequent conductance access, we could achieve >90% accuracy even in high NL and low AF cases ($|NL(LTD)| \geq 8$ or $|NL(LTP)| \geq 8$ except for $|NL(LTP)|=|NL(LTD)|=9$). The proposed weight update method is expected to increase the performance of a network consisting of highly nonlinear synaptic devices in a hardware-efficient manner.

### THE CONCEPT OF UPDATE NOISE AND METRICS USED TO EVALUATE UPDATE NOISE

In this work, one of the main causes of accuracy degradation in HNN is update noise due to the nonlinear conductance update characteristics of synaptic devices. We use the following equation to quantitatively evaluate the UN (Update Noise) generated in a training iteration:

$$\text{UN}(W_n) = ||P_n k - (W_{n,\text{updated}} - W_n)|| \quad (L2\ norm) \quad (5)$$

where $k$, $P_n$, $W_n$, and $W_{n,updated}$ represent the scaling factor that scales the number of applied pulses into weight scale, a vector of the number of pulses applied to synaptic devices, a weight vector before weight update, and the weight vector after weight update at $n^{\text{th}}$ training iteration, respectively.

Software-based studies have shown that gradient noises with certain structures have regularization effects [29]. However, to the best of our knowledge, the structure and the effect of the gradient in HNN are not well understood. Moreover, for on-chip training, it is difficult to freely manipulate the gradient in the desired way. Thus, we try to self-adaptively reduce the average amount of UN evaluated after training. We use the following equation to evaluate AUN (Average Update Noise) after the end of a training case:

$$\text{AUN}(W) = \sum_{n, ||P_n|| \neq 0} \frac{1}{m} ||P_n k - (W_{n,\text{updated}} - W_n)|| \quad (L2\ norm) \quad (6)$$

where $L$, $n$, and $m$ represent the loss function of a network, the total number of training iterations of an algorithm, and the total number of training iterations such that the sum of the number of applied pulses in all synaptic devices at a given iteration is not equal to zero, respectively.

### PROPOSED WEIGHT UPDATE SCHEME

To tackle the limitations in the existing non-linearity mitigation techniques, we propose a new weight update scheme.

We analyze our proposed weight update rule through a simulator named NeuroSim using the Modified National Institute of Standards and Technology (MNIST) handwritten digit database [30]. The model comprises three fully connected layers where the numbers of input layer neurons, hidden layer neurons, and final layer neurons are 400, 100, and 10, respectively. The activation function that we use is the sigmoid function. The input image pixel precision is 1 bit and the precision of neurons in the hidden layer and the output layer is 8 bit. The weights are initialized by randomly choosing a number from [-1, $-\frac{2}{3}$, $-\frac{1}{3}$, 0, $\frac{1}{3}$, $\frac{2}{3}$, 1]. A set of 60000 images is used for training and a set of 10000 images is used for the test. Each of the simulation cases is trained up to 100 epochs based on the Stochastic Gradient Descent (SGD) algorithm. Networks under different hyperparameter and device non-linearity settings have different optimized numbers of training epochs to yield a high level of test accuracy. Thus, the accuracy we report is the statistical average of the highest value among test accuracies evaluated after each epoch.

We first consider a synaptic device that satisfies NL(LTP)=1, NL(LTD)=-9, and $P_{max}$=100. By setting NL(LTD)= -9, the AF of LTD becomes 3 and the conductance falls to 0.8% of its maximum conductance within 11 pulses as seen from Fig. 3(a)-(b). This setting can be used to simulate any synaptic devices where LTD non-linearity or AF must be substantially sacrificed to enhance other crucial device properties for HNN training. Starting from this case, we will show that our proposed method can be applied to other highly nonlinear NL cases (|NL(LTD)|≥ 8 or |NL(LTP)| ≥8 but not |NL(LTP)|=|NL(LTD)|=9). We only consider the cases where NL(LTP) and NL(LTD) have different polarities.

We now introduce a novel weight update method named conditional reverse update scheme (CRUS) that can effectively control the AUN of an algorithm. The main goal of this work is to design a weight update method that (i) ultimately reduces AUN ($W_{HO}$) evaluated at the end of the training with infrequent and inaccurate conductance sensing and (ii) does not require any pulse shape or duration tuning. Note that AUN ($W_{IH}$) can be evaluated as well. However, it will be shown that AUN ($W_{HO}$) is the most influential factor determining the accuracy of the network.

In this scheme, differential synapse pairs are used to represent each weight element. For notational simplicity, we refer to synapse pairs as synapses. Each of the synapses consists of two bidirectional synaptic devices with conductance values $G_p$ and $G_n$. The weight value $W$ that each synapse represents can be expressed as $W = \frac{G_p - G_n}{G_{max} - G_{min}}$. From this point, we define the synapses located between the hidden and the output layer as $S_{HO}$ and the corresponding weight matrix as $W_{HO}$. Similarly, synapses located between the hidden and the input layer and the corresponding weight matrix are $S_{IH}$ and $W_{HO}$, respectively.

For hardware simplicity, we separate the weight update process into two phases, and training iteration corresponds to either the normal update phase or the reverse update phase. When a specific training iteration corresponds to the normal update phase, only LTP updates are used for conductance updates. In contrast, only LTD updates are used for conductance updates in the reverse update phase. In addition, the reverse update phase occurs only once every reverse update period. The reverse update period can range from 2 to 10 depending on the hyperparameters of the algorithm or the type of synaptic device used in HNN. By allowing both LTP and LTD updates, we can prevent the saturation of conductance or the generation of stuck-at-fault synapses which degrade the performance of the network significantly [31]. The separation of the training into two phases also allows us to use separate learning rates for the LTP and LTD updates. This strategy can compensate for the inherent asymmetrical update characteristics of synaptic devices [32] parallelly.

To effectively reduce AUN, we apply an additional condition during the reverse update phase. The key idea of the conditional reverse update scheme is to conditionally skip LTD updates during the reverse update phase. The decision of whether to skip the LTD update is determined by the stored bit designated for each synaptic device in HNN. The stored bits are set to 1 if the conductance values of their corresponding synaptic devices are lower than $G_{th}$, and to 0, otherwise. In addition, the stored bits are periodically refreshed. If an algorithm calls for a weight update in $G_p^*$ ($G_n^*$) during the reverse update phase, we do not apply any pulses to $G_p^*$ ($G_n^*$) if the stored bit, that is designated for its complementary device, $G_n^*$ ($G_p^*$) is 1. By adjusting $G_{th}$, we can (i) control UN caused by synapses with certain conductance configurations and (ii) self-adaptively induce synapses to have desirable conductance configurations during training. Therefore, by setting a proper $G_{th}$, we can effectively reduce AUN. For better understanding, we present pseudocode that demonstrates our proposed update method below.

---

**[Proposed Weight Update Algorithm]**

$W = \dfrac{G_p - G_n}{G_{max} - G_{min}}$;

Let $m$ be the total number of training iterations;
Let $rp$ be the reverse update period;
Let $refp$ be the reference period;
Let $P_i$ be the number of pulses applied at training iteration $i$;
Let $a_n$ be the learning rate for the normal update phase;
Let $a_r$ be the learning rate for the reverse update phase;
**for** $i = 1$ to $m$ **do**
   calculate $\dfrac{dL}{dW}$ ;
  **if** $i$ is 1 or integer multiple of $refp$ **then**
    store (bool) $flag\_p = (G_n < G_{th})$;
    store (bool) $flag\_n = (G_p < G_{th})$;
  **endif**
  **if** $i$ is integer multiple of $rp$ **then**
    Convert $|a_r \dfrac{dL}{dW}|$ to $P_i$;
    **if** $\left(\dfrac{dL}{dW} > 0\right)$ **then**
     Apply $P_i$ LTD pulses to $G_n$ *if not flag_n*;
    **else**
     Apply $P_i$ LTD pulses to $G_p$ *if not flag_p*;
    **endif**
  **else**
    Convert $|a_n \dfrac{dL}{dW}|$ to $P_i$;
    **if** $\left(\dfrac{dL}{dW} > 0\right)$ **then**
     Apply $P_i$ LTP pulses to $G_p$;
    **else**
     Apply $P_i$ LTP pulses to $G_n$;
    **endif**
**endfor**

---

For convenience, we only present the pseudocode for |NL(LTP)| ≤ |NL(LTD)| cases. |NL(LTP)| ≥ |NL(LTD)| cases can be handled by changing the LTD skip condition and swapping the definition of the reverse update phase and the normal update phase. The rest of this paper explains the |NL(LTP)| ≤ |NL(LTD)| case. We will show that the reference period can be made longer than the optimal refresh period.

We first see how CRUS effectively reduces the UN of individual synapses. Here, the UN of a synapse is defined as UN($W$), where $W$ represents the one-element weight vector corresponding to the synapse. By reducing the UN of individual synapses, we can reduce UN($W_{IH}$) and UN($W_{HO}$) at each training iteration, and thereby minimize AUN($W_{IH}$) and AUN($W_{HO}$). To measure the amount of UN reduction, we calculate the expected UN of an individual synapse when the reverse period is set to 2. Here, not considering the actual training situation, we assume that the pulses to be applied during each weight update are extracted from a uniform [-3, 3] distribution and the reverse update phase occurs with a probability of 50%. To see how the expected UN varies with the conductance configuration ($G_p$, $G_n$), we also characterize the synapses based on their conductance configurations as follows:

(0,0) synapse: $G_p < G_{th}, G_n < G_{th}$
(0,1) synapse: $G_p < G_{th}, G_n \geq G_{th}$
(1,0) synapse: $G_p \geq G_{th}, G_n < G_{th}$
(1,1) synapse: $G_p \geq G_{th}, G_n \geq G_{th}$

Fig. 5(a)-(e) show the initialized conductance configuration distribution and the expected UN of synapses with different conductance configurations under different $G_{th}$ values. It can be seen that the expected UNs of the (0,0), (0,1), and (1,0) synapses are relatively low compared to those of (1,1) synapses. This is because the LTD skip condition applied to (0,0), (0,1), and (1,0) synapses effectively suppresses the expected UN. This implies that the AUN can be significantly reduced when the synapses mostly stay as (0,0), (0,1), (1,1) types during training. Thus, it is important to keep the number of (1,1) synapses as low as possible to minimize AUN.

Now, we discuss how to decrease the number of (1,1) synapses during training. As training proceeds, there are frequent conversions between synapse types. The conversion from (0,1) or (1,0) synapses to (1,1) synapses should be minimized to suppress the number of (1,1) synapses during training. According to previous work, in a network composed of bidirectional synaptic devices, the conductance values tend to converge around the "symmetry point" [33]. This is the conductance value $G_{sym}$ that satisfies $|\frac{dG_{LTP}(G^{-1}_{LTP}(G_{sym}))}{dP}| = |\frac{dG_{LTD}(G^{-1}_{LTD}(G_{sym}))}{dP}|$ (here, $G_{LTP}(P)$ and $G_{LTD}(P)$ are the functions used in Eqs. (1) and (2), respectively). Note that this is also the point to which the conductance value eventually converges after one LTP pulse and one LTD pulse are continually applied to a synaptic device in an alternating fashion. A similar phenomenon can be seen under CRUS as well. From Fig. 6, it can be seen that the conductance distribution of $G_p$ in (0,1) synapses or $G_n$ in (1,0) synapses tends to converge around the symmetry point (~0.32 for the NL(LTP)=1, NL(LTD)=-9 case). Thus, as $G_{th}$ increases above the symmetry point, the average distance between $G_{th}$ and the conductance values of the (0,1) and (1,0) synapses increase. As a result, it becomes more unlikely that the conductance values of such synapses cross over $G_{th}$ during training. From Fig. 7 (a), it can be seen that the number of (1,1) synapse types decreases as $G_{th}$ increases, which is aligned with our analysis. Therefore, the key strategy to reduce the number of (1,1) synapses is to increase $G_{th}$.

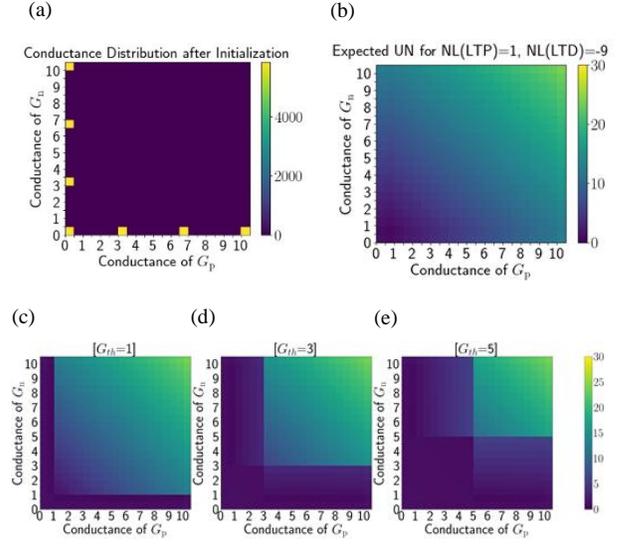

Fig. 5. (a) Synapse distribution after initialization. (b) Expected UN without LTD skip condition ($G_{th}$=0). Expected UN for (c) $G_{th}$=1 (d) $G_{th}$=3 (e) $G_{th}$=5.

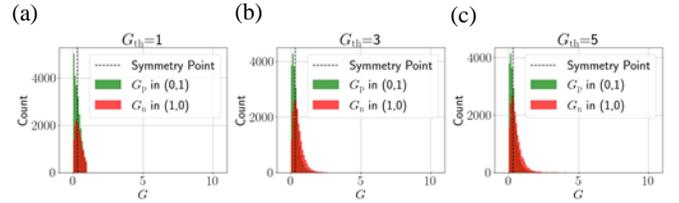

Fig. 6. Conductance distribution of $G_p$ in (0,1) synapses and $G_n$ in (1,0) synapses. (a) $G_{th}$=1, (b) $G_{th}$=3, (c) $G_{th}$=5 case. The conductance values converge around the symmetry point regardless of the $G_{th}$ value.

Finally, we analyze how increasing $G_{th}$ affects AUN($W_{IH}$) and AUN($W_{HO}$). We then determine how we can use AUN($W_{IH}$) and AUN($W_{HO}$) to optimize $G_{th}$ in terms of accuracy. To determine the AUN component of a specific synapse type, we extend the definition of AUN hereafter. AUN($W^{synapse\ type}$) is calculated by masking the elements that do not correspond to the superscripted synapse type for $P_n$ and $W_n$ in the argument of the summation in Eqs. (6).

As shown from Fig. 7(b)-(c), increasing $G_{th}$ does not always lead to a decrease in AUN($W_{IH}$) and AUN($W_{HO}$) and an increase in accuracy. Several important factors should be considered simultaneously to fully understand the unexpected behaviors of AUN($W_{IH}$), AUN($W_{HO}$), and accuracy with respect to $G_{th}$ and optimize $G_{th}$ in terms of accuracy.

First, the effect of the LTD skip condition decreases as $G_{th}$ increases. This arises from the difference in the ratio of non-zero pulses in the reverse update phase with respect to $G_{th}$. It is important to note that for (0,1) synapses, only LTD pulses applied to $G_n$ are skipped, and for (1,0) synapses, only LTD pulses applied to $G_p$ are skipped. For this reason, it is possible that depending on the pulse input patterns, the effect of the LTD skip condition may vary. It can be seen from Fig. 8 that the average ratio of non-zero pulses applied to (0,1) and (1,0) synapses in the reverse update phase increases as $G_{th}$ increases. Therefore, combined with the effect of the increase in the number of (0,1) and (1,0) synapses with a $G_{th}$ increase, AUN($W_{HO}$) and AUN($W_{IH}$) can increase as $G_{th}$ increases. This implies that $G_{th}$ should not be increased above the point where the increase in AUN($W_{IH}^{(0,1)}$), AUN($W_{IH}^{(1,0)}$), AUN($W_{HO}^{(0,1)}$), and AUN($W_{HO}^{(1,0)}$) starts to offset the decrease in AUN($W_{IH}^{(1,1)}$), and AUN($W_{HO}^{(1,1)}$).

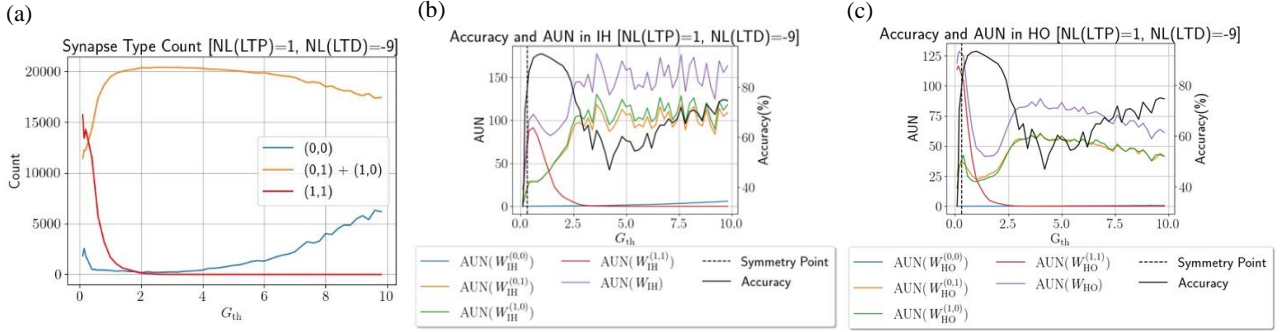

Fig. 7. Plot of (a) Number of each synapse type averaged over the entire training process. (b) AUN($W_{IH}$), AUN($W_{IH}^{(0,0)}$), AUN($W_{IH}^{(0,1)}$), AUN($W_{IH}^{(1,0)}$), AUN($W_{IH}^{(1,1)}$). (c) AUN($W_{HO}$), AUN($W_{HO}^{(0,0)}$), AUN($W_{HO}^{(1,0)}$), AUN($W_{HO}^{(0,1)}$), AUN($W_{HO}^{(1,1)}$). Here, the AUN values are scaled by a factor of 10000.

In addition, there is a different behavior of AUN and accuracy with respect to $G_{th}$ when $G_{th}$ is lower than the symmetry point. It can be seen from Fig. 7(b)-(c) that if $G_{th}$ is low ($G_{th}$<0.32), both AUN($W_{IH}$) and accuracy increase with increasing $G_{th}$. This differs from the case of AUN($W_{HO}$), which generally has a contrary relationship with accuracy with respect to $G_{th}$. This is because if $G_{th}$ is set near or lower than the symmetry point (~0.32), the $G_p$s and $G_n$s of (1,1) synapses in $S_{HO}$ gradually become densely populated around the symmetry point (see Fig. 9(a)), causing excessive UN in $W_{HO}$. If the symmetry point is located above $G_{th}$, once a synapse changes to (1,1), it rarely switches to (0,1) or (1,0). In this case, $G_p$ and $G_n$ keep converging around the symmetry point until the end of training. Consequently, the weight value that the synapse represents becomes increasingly smaller as training proceeds. This leads to an excessive increase in the number of low magnitude weights in $W_{HO}$ at low $G_{th}$ ($G_{th}$<0.32).

Similar to a reported observation that excessive sparsity of weights leads to vanishing gradient problem [34], excessive increase in the low magnitude weights in $W_{HO}$ limits gradient flow to $W_{IH}$ at low $G_{th}$. The vanishing gradient problem retards training in early layers and significantly drops the accuracy of a network [35]. From Fig. 9(b), it can be seen that the gradient passed to $W_{IH}$ significantly decreases as $G_{th}$ decreases. Referring to Fig. 5, this implies that the synapses corresponding to $W_{IH}$ will stay at the low expected UN region throughout training. This explains why decreasing $G_{th}$ below the symmetry point can lead to a drop in both AUN($W_{IH}$) and accuracy. Therefore, $G_{th}$ should be set higher than the symmetry point for accuracy enhancement. This also implies that lower AUN($W_{IH}$) does not guarantee higher accuracy. This is because the magnitude of the gradient passed to $W_{IH}$ is the limiting factor of accuracy at low $G_{th}$ ($G_{th}$<0.32). If $G_{th}$ increases toward the symmetry point, $W_{IH}$ gradually receives more gradients as seen from Fig. 9 (b). AUN($W_{IH}$) increases with the $G_{th}$ increase until the decreasing number of (1,1) synapses starts to have a dominant impact on AUN($W_{IH}$). From this point, AUN($W_{IH}$) and AUN($W_{HO}$) have similar trends with respect to $G_{th}$. This explains the difference in the behavior of AUN($W_{HO}$) and AUN($W_{IH}$) with respect to $G_{th}$. Therefore, it can be concluded that AUN($W_{HO}$) is the most relevant factor that should be considered for accuracy optimization. Fig. 10 shows the correlation between the $G_{th}$ that minimizes AUN($W_{HO}$) and optimized $G_{th}$ in terms of accuracy under different NL settings. This shows that by fine-tuning $G_{th}$ around the value that yields minimal AUN($W_{HO}$), we can easily optimize a network in terms of accuracy.

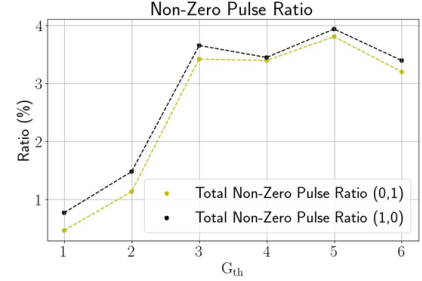

Fig. 8. Ratio of non-zero pulses applied to (0,1) and (1,0) synapses in reverse update phase with respect to $G_{th}$.

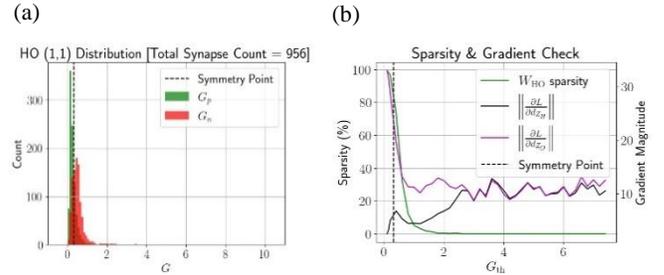

Fig. 9. (a) $G_p$ and $G_n$ distribution of (1,1) synapses located between the hidden and the output layer at low $G_{th}$ ($G_{th}$<0.32). (b) Derivative of loss function($L$) with respect to hidden layer pre-activations($z_H$) and output layer pre-activations($z_O$) averaged over the whole training process and sparsity. Here, sparsity is defined by the ratio of weights of magnitude <0.2.

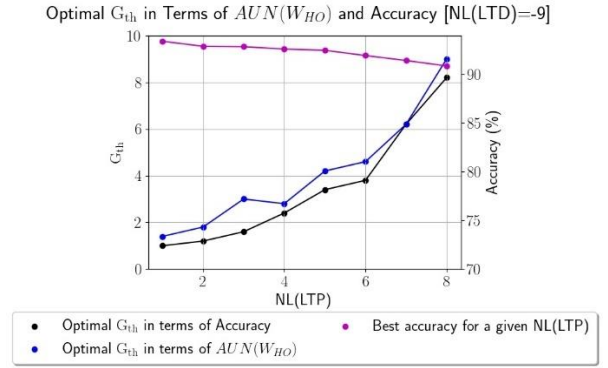

Fig. 10. Correlation between optimal $G_{th}$ in terms of AUN($W_{HO}$) and that in terms of the network's accuracy.

### EXPERIMENTAL RESULTS AND DISCUSSION

*A. Results with various device non-linearity and hyperparameter conditions*

In this section, we show that CRUS can be applied to other NL and hyperparameter settings. As discussed earlier, suppressing the generation of (1,1) synapses and addressing the vanishing gradient problem by increasing $G_{th}$ above the symmetry point is the key to minimizing AUN($W_{HO}$). Meanwhile, too large a $G_{th}$ leads to an increase in AUN($W_{HO}$) and degrades accuracy due to the UN generated by (0,1) and (1,0) synapses. Therefore, if there is a shift in the symmetry point, the conductance distribution of $G_p$ in the (0,1) synapse, or the conductance distribution of $G_n$ in the (1,0) synapse, $G_{th}$ needs to be adjusted accordingly. If NL and hyperparameters of the network are chosen to shift these three factors upward, $G_{th}$ should be increased to suppress $G_{th}$ crossover events. In the opposite case, $G_{th}$ should be decreased to reduce the effect of increasing AUN($W_{HO}^{(0,1)}$), and AUN($W_{HO}^{(1,0)}$).

First, we analyze the impact of NL in choosing optimal $G_{th}$. In this section, the refresh period, reference period, $P_{max}$ are set to 2, 2, 100, respectively. Increasing NL(LTP) is expected to increase the optimal $G_{th}$ since the symmetric point is shifted upward in this case. Similarly, decreasing NL(LTD) is expected to have the opposite effect. The simulation results shown in Fig. 11(a)-(b) verify the expected tendencies.

A change in the ratio of the learning rate corresponding to the normal update phase ($\alpha_n$) and that corresponding to the reverse update phase ($\alpha_r$) also affects the choice of $G_{th}$. For instance, increasing $\alpha_n$ while maintaining $\alpha_r$ shifts the conductance distribution of $G_p$ in the (0,1) synapses or $G_n$ in the (1,0) synapses upward. This is because the conductance generally increases at a faster rate than that when $\alpha_n=\alpha_r$. Thus, as $\alpha_n$ increases, $G_{th}$ needs to be increased for accuracy enhancement. Using this fact, we tested various values for $\alpha_n$ and $\alpha_r$ to further optimize the accuracy. From the results shown in Fig. 11(a)-(b) it can be seen that CRUS yields >90% accuracy for all NL configurations in which |NL(LTP)| ≥ 8 or |NL(LTD)| ≥ 8 (except for the NL(LTP)=9 and NL(LTD)=-9 case) through setting an optimized $G_{th}$ value for given $\alpha_n, \alpha_r$, NL(LTP), and NL(LTD). In addition, CRUS yields better accuracy than the refresh method if NL(LTP) exceeds 1.

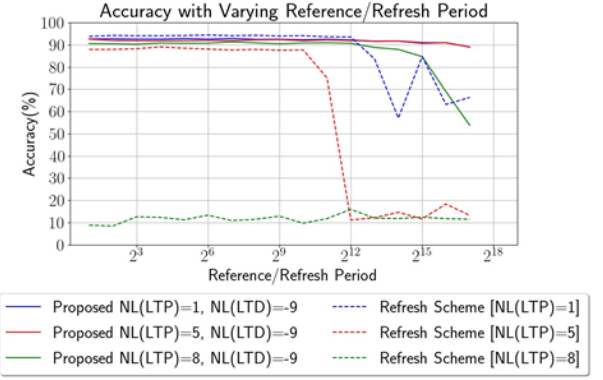

Fig. 12. Three exemplary cases of CRUS to demonstrate its robustness to infrequent conductance sensing. Reverse update period is abbreviated as "rup".

Extending the reverse update period has a similar effect with respect to increasing $\alpha_n$ since LTP updates happen more frequently with increasing reverse update period. Extending the reverse update period can offer hardware advantages when the LTD process is slower than the LTP process or consumes more power than it [36,37]. By choosing an optimal reverse update period, LTD update events can be reduced and lead to improvements in latency and power consumption. The simulation results displayed in Fig. 11(c) show that the optimal $G_{th}$ increases with the increase in the reverse update period. However, the accuracy tends to decrease when the reverse update period increases. Therefore, latency, power, and accuracy tradeoff should be considered when adjusting the reverse update period.

*B. Effect of the reference period on the training performance and comparison with the refresh method*

The key advantage of CRUS is that conductance can be more infrequently sensed than the refresh method. This is true regardless of the choice of NL and the reverse update period. It can be seen from Fig. 12 that CRUS generally shows better accuracy under a longer conductance access period and the accuracy enhancement increases as NL(LTP) increases. Infrequent and inaccurate conductance sensing is expected to accelerate the speed of training while providing advantages in peripheral circuit cost, memory cost, and chip size. Accordingly, CRUS is expected to provide excellent hardware efficiency in HNN operation.

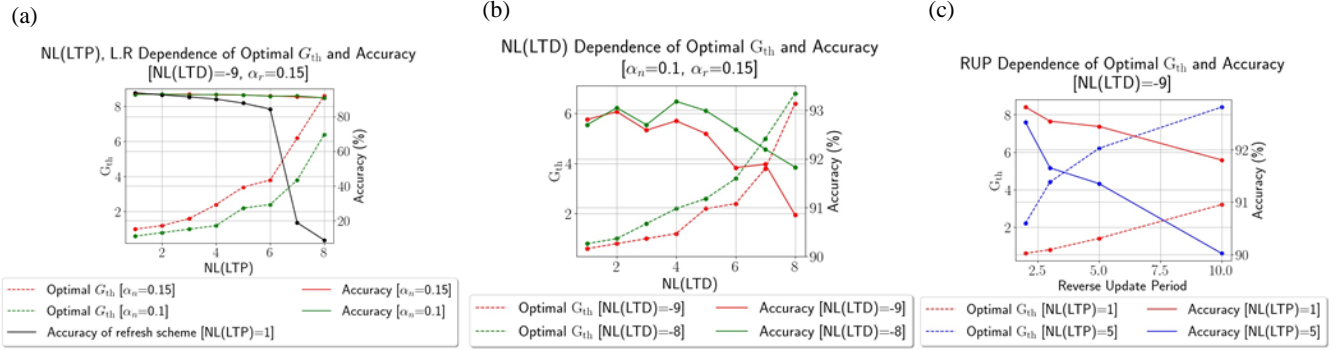

Fig. 11. Accuracy and optimized $G_{th}$ under various hyperparameter, NL(LTP), NL(LTD) conditions: (a) NL(LTP), $\alpha_n$ dependence of accuracy and optimized $G_{th}$, (b) NL(LTD) dependence of accuracy and optimized $G_{th}$, (c) reverse update period dependence of accuracy and optimized $G_{th}$.

### COMPARISON WITH OTHER NON-LINEARITY MITIGATION SCHEMES

Table 1 compares the accuracy of the network under CRUS against previously reported non-linearity mitigation methods. For a fair comparison of the network performance, we set hardware conditions equal to or lower than that of the previous works. We also perform additional simulations on the PL method to test its effectiveness under severe NL cases. Note that ANL denotes a metric proposed to evaluate the asymmetry between LTP and LTD curves [28]. 0 is the case of ideal symmetry, and the asymmetry increases as the ANL value increases. The input image precision is increased to fairly compare the results with those of the prior work which assumes 256 gray levels of the input pixels. Interestingly, our method even works under a lower number of conductance states (=64) and still achieves >89% accuracy if the input pixel precision is increased. From these results, it can be concluded that CRUS shows superior accuracy and hardware performance compared to previous non-linearity mitigation techniques.

### ROBUSTNESS TO TEMPORAL VARIATIONS IN CONDUCTANCE UPDATE

Temporal variation in the conductance update process is another crucial factor to be considered in designing HNN algorithms [38,39]. Therefore, in this section, we analyze CRUS under temporal variation in conductance updates and demonstrate its robustness to such effects. The equations used to model the temporal variation at every LTP and LTD update event are as follows [19]:

$G_{i,\text{updated}} = G_i + Z\sqrt{P_i}, Z \sim N(0, \sigma^2)$ (7)

$\sigma = \alpha(G_{\max} - G_{\min})$ (8)

where $G_i$, $G_{i,\text{updated}}$, $N(0, \sigma^2)$, $P_i$ and $\alpha$ are the conductance value of a synaptic device at $i^{\text{th}}$ training iteration before weight update, the conductance value of the same device at $i^{\text{th}}$ training iteration after weight update, the normal distribution with mean 0 and standard deviation $\sigma$, the number of applied pulses at $i^{\text{th}}$ training iteration, and the variation coefficient respectively. It can be seen from Fig. 13 that CRUS is highly resilient to temporal conductance variation compared to the refresh method and one ideal synaptic device (a device where both LTP and LTD are perfectly linear) per synapse method. This is because frequent LTD skip update events help minimize the damage caused by temporal variations in conductance updates.

### CONCLUSION

In this paper, CRUS has been proposed to mitigate highly nonlinear or abrupt conductance update characteristics of synaptic devices. This method has shown high accuracy (>90 %) on the MNIST classification task under $P_{\max}=100$, $|\text{NL(LTP)}| \geq 8$ or $|\text{NL(LTD)}| \geq 8$ cases (except for the NL(LTP)=9 and NL(LTD)=-9 case) by high hardware efficiency. Analysis of the training dynamics of the network using UN has implied the correlation between the optimal $G_{th}$ in terms of AUN and the optimal $G_{th}$ in terms of accuracy. Based on the change in the symmetry point under various situations, we have introduced methodologies to optimize $G_{th}$ depending on NL(LTP), NL(LTD), learning rate, and the reverse update period. These methodologies are expected to provide additional flexibility in hardware design and lead to training performance enhancement. Finally, we have shown that CRUS exhibits robustness to cycle-to-cycle variation in conductance updates. We believe the proposed weight update scheme, update noise evaluation metric, and the methodologies for $G_{th}$ tuning can serve as a useful tool when designing and optimizing HNN consisting of synapses with poor non-linearity conditions. They would be more appealing if they can be (i) tested to large-scale networks, various activation function settings, and different domain problems, and (ii) a parallel weight update scheme to sense conductance in 1-bit and skip update is provided. This will be the future direction of our work.

| Weight Update Method | Weight Precision (Levels) | Input Image Pixel Precision (Levels) | Neuron Precision (Levels) | Non-linearity Condition | Accuracy (%) |
|---|---|---|---|---|---|
| [21] (4 segment division for both LTP and LTD curve) | 100 | 2 | 256 | NL(LTP)=8 NL(LTD)=-8 | 66.98 |
| | | | | NL(LTP)=1 NL(LTD)=-9 | 71.9 |
| **Our method [Reference period=4096]** | 100 | 2 | 256 | NL(LTP)=8 NL(LTD)=-8 | 91.14 |
| | | | | NL(LTP)=1 NL(LTD)=-9 | 92.41 |
| [28] | 64 | 256 | 256 | ANL=0.8 | 87.8 |
| **Our Method [Reference Period=4096]** | 64 | 4 | 256 | NL(LTP)=8 NL(LTD)=-8 [ANL=0.99] | 89.66 |
| **Our method [Reference period=4096]** | 64 | 4 | 256 | NL(LTP)=4 NL(LTD)=-9 [ANL=0.86] | 92.89 |

Table 1. Comparison of accuracies achieved by CRUS and those achieved by previously reported non-linear LTP and LTD mitigation techniques.

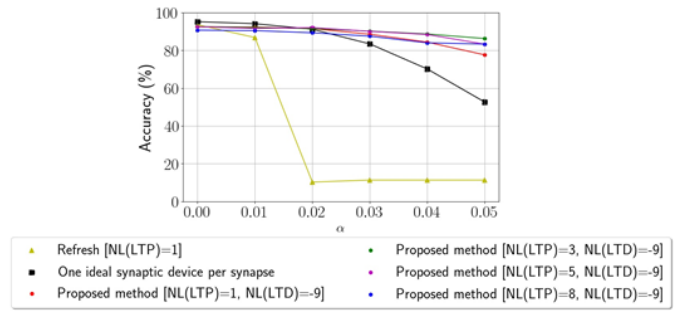

Fig. 13. Comparison of robustness to temporal variations in conductance update between different methods. Here, NL(LTD)=-9, and the reference period and the reverse update period are set to 4096 and 2, respectively. The refresh period for the refresh scheme is set to 2048.